\documentclass{appolb}
\usepackage{epsfig}
\begin{document}
\title{
The average behaviour  of financial market \\ 
by 2 scale homogenisation
%The Paper Title comes Here...%
%\thanks{Presented at ...}%
}
\author{
R. Wojnar
%Put here the name(s) of the Author(s)
\address%{and affiliation}
ul. {\'S}wi\c{e}tokrzyska 21, 00-049 Warszawa, \\ 
IPPT, Polska Akademia Nauk
%\and
%the Name(s) of other Author(s)
%\address{and their affiliation}
}
\maketitle
\begin{abstract}
The financial market is nonpredictable, as according to the Bachelier, the
mathematical expectation of the speculator is zero. Nevertheless, we observe in
the price fluctuations the two distinct scales, short and long time. Behaviour
of a market in long terms, such as year intervals, is different from that in
short terms.
  
 A diffusion equation with a time dependent diffusion coefficient that
describes the fluctuations of the financial market, is subject to a two-scale
homogenisation, and long term characteristics of the market such as mean
behaviour of price and variance, are obtained.
We indicate also that introduction of convolution into diffusion equation permits to obtain L- stable behaviour of finance. 

%Here comes the abstract
\end{abstract}
  
\section{Introduction}

The prices on stock  market are formed as a result of superposition of large number of different reasons and can be assumed to be governed by probability laws.  
The fluctuations of prices on stock market resemble an errating walk, as it was indicated yet in 1900 by Louis Bachelier [1], when he derived the diffusion equation from a condition that speculators should receive no information from the past prices.   The difference of  action prices 
$ 
x = x(t)  \equiv p(t + \Delta t) - p(t) 
$,
observed at two  time moments $ t $  and $ t + \Delta t$, 
plays in this  diffusion equation role of independent spatial variable. 
Hence,   the motion of prices on the financial market is similar to the brownian movement, discovered by the biologiste  Robert Brown [2] and analysed  by Albert Einstein [3,4] and Marian Smoluchowski [5-7], cf. also [8]. Bachelier's observation did not find large recognition at his life, but now is a basis of greater part of modeles of prices, especially the Black-Scholes model [9], cf. also [10].
Later  Paul A. Samuelson [11] indicated that instead of a simple difference (1)  it is more proper to  consider  differences of  the respective logarithms 
$
x = \ln p( t + \Delta t) - \ln p( t ) 
$.  

However, as it was indicated by Benoit Mandelbrot [12], despite the fundamental importance of Bachelier's random walk of the price changes (one cannot imagine an advanced textbook on  finances without the  brownian motion description as its starting point), 
the empirical samples  of  successive differences of stock price changes  gathered from 1890 year,  are not normally distributed: they  are usually too peaked to be  Gaussian and  do not have finite variance.  The distribution of price changes is leptokurtic, since the sample kurtosis is much greater than 3, the value for a normal distribution.
Mandelbrot regarded that the  price changes belong to the stable family of distributions, known as L-stable or L\'evy-Pareto distributions. Mandelbrot and Wallis [13] distinguished  two non-Gaussian kinds of events  observed in the economic world: isolated catastrophic events, the Noah effect which refers to abrupt and discontinuous changes in speculative time series and regular alternations of good and bad series, termed the Joseph effect.

Besides those effects with stochastic non-gaussian origin,  
another type of departure from normal distribution is observed when
the irregular random behaviour of stock price changes is superposed on another  regular periodic pattern. 
There is a definite evidence of periodic  behaviour of price changes corresponding to intervals of a day, week, quarter and year, according to the rhytm of human activity. 
Maury Osborne [14] indicates, for example, that there is a reproducible burst of trading at the beginning and the end of trading day. While diurnal cycle is almost obvious, a somewhat more subtle statistical analysis ($\chi^2$ test) reveals  a week periodicity in the daily across-the-market dispersion of stock price changes. This price dispersion is a maximum at middle of week, what  can be interpreted that traders tend to forget the market business over a long week-end and make up their minds at the beginning of new week.     

At the beginning of the present paper, we  outline some properties  of 
diffusion equation with nonhomogeneous coefficient (dependent on time $t$ or price changes $x$) and   describe 
its solutions as  the Gauss and L\'evy types. 
We also propose to use a two scale homogenisation
method to describe an average behaviour of a financial market in a long time or in averaged market
in the case in which the diffusion coefficient depends on stock price change.

\section{Diffusion in 1 dimension}

Movement  of brownian particle  is described by a distribution function $f=f(x,t)$  satisfying the diffusion equation 
\begin{equation}%1
\frac{\partial f}{\partial t} = \frac{\partial }{\partial x}\left( D  \frac{\partial f}{\partial x}\right)
\end{equation}
Function $f$ gives the  probability density of finding Brownian particle at position $x$ at time $t$, and $D$ denotes the coefficient of diffusion.
According to the Einstein  
fluctuational dissipative relation 
$ 
D \sim T / \eta 
$, 
the coefficient $D$ is proportional to a quotient of  the absolute temperature $T$ and viscosity $\eta$, and if $\eta$ does not depend on $T$, it is simply proportional to $T$. 

The form of Eq.(1) admits dependence of the coefficient $D$ on $x$ which may be realized e.g. by dependence of $T$ on $x$.  If  $D$ is a function of time $D=D(t)$ only, or if it  is constant, the following  form  is obtained
\begin{equation}%2
\frac{\partial f}{\partial t} = D  \frac{\partial^2 f}{\partial x^2}
\end{equation}
Depending on interpretation, the coefficient $D$ denotes either the thermal diffusivity (quotient of the heat condictivity and proper heat) or the diffusion coefficient. The last meaning is used below.

According to Bachelier's observation, the  time independent variable $t$ in diffusion equation is measured by successive nuber of stock transactions and the independent variable $x$,  denotes stock action price change.   The coefficient $D$ varies according to a market temperature, cf. [15].

\subsection{Fick's equations}
Let $f=f(x, t)$ be the probability  density  of   finding  
a  brownian (B.)  particle  at point  $ x $  and at  time  $t$, and let
$j=j(x, t)$ be  a  stream  of   B. particles.
The continuity (or balance)  equation describes conservation of the number of  B. particles
\begin{equation}%3
{\partial f \over \partial t} + {\partial j \over \partial x}   =  0
\end{equation}
The transport relation, known as the first law of Fick reads
\begin{equation} %4
 j  =  - D   {\partial f \over \partial x} 
\end{equation} 
where $ D $  denotes the  diffusion coefficient, and we admit in general   $ D=D(x, t) $. 
The first Fick's law  
extended for the case of presence of external forces $ F $ has the form
\begin{equation} %5
 j  =  - D  \left( {\partial f \over \partial x} -  {F \over T}  f  \right) 
\end{equation}
where $ T $  is a  temperature.
From the mass balance and the first Fick law, the second Fick law - it is the diffusion equation (1) - can be derived.
 
\subsection{Steady diffusion in temperature gradient}

Let a  diffusion in a slab $ 0 \le x \le L$  be stationary
$ 
j = J_0 
$ =  constant. 
In presence of an external force $F$, when the concentration within the diffusion volume does not change with respect to time 
($j=$ constant), 
the Fick first law has a form 
$$
  - D  \left( {\partial f \over \partial x} -  {F \over T}  f  \right) = J_0 
$$
In special case, when $J_0=0$ and   the ends of the slab are kept at different temperatures
$
T(x=0)=T_0,   T(x=L)=T_L, 
$
what gives a linear temperature distribution
$
T = A x + B  
$, 
we obtain 
\begin{equation} %6
f = C (A x + B)^{F/A}  
\end{equation} 
Here
$
A= (T_L-T_0) / L$  and $ B=T_0$,  
while the constant $ C $  normalizes the distribution 
$
  \int_0^L f dx = 1
$.
We observe that even in such simple case the distribution $f$ in slab  is no longer gibbsian.

\section{Time dependent coefficient of diffusion }

In this case the diffusion equation has the form (2). If  $f(x, 0) =  \delta(x)$, the solution of (2) is, cf. [16], 
\begin{equation} %7
f(x,t) = \frac{1}{2 \sqrt{ \pi \,  \int_0^t Ddt}} \ {\rm e}^{- \, \frac{x^2}{4 \int_0^t Ddt}}
\end{equation} 
The variance of this distribution is 
\begin{equation} %8
\sigma^2 = \sigma^2(t) \equiv 2  \int_0^t D(\tau) d\tau
\end{equation} 
Hence
\begin{equation} %9
f(x,t) = \frac{1}{\sqrt{2 \pi} \, \sigma} \ {\rm e}^{- \, \frac{x^2}{2 \sigma^2}}
\end{equation} 
If  the diffusion coefficient $D$ does not depend on $ t $ and is constant, we have for the dispersion  (standard deviation)
\begin{equation} %10
\sigma =  \sqrt{2  D t} 
\end{equation} 
the classical  result for the gaussian distribution in one-dimensional process.

\subsection{Periodic time dependence of  the diffusion coefficient}

As it was said it is  observed a periodic oscillation of the across-the-market dispersion of price change  for time intervals (day, week, and so on). As the dispersion is proportional to the diffusion coefficient $D$, it means that $D$ is a periodic function of time. 

Therefore,  let $D(t)$  be a function of time with  period $T$.  
For   $t = n T $, with a whole number $n$,  we have
\begin{equation} %11
\int_0^t D dt =  \int_0^T D dt +  \int_T^{2T} D dt + \cdots + \int_{(n-1) T}^{n T} D dt = n  \int_0^T D dt 
\end{equation} 
Hence, according to (8)
\begin{equation} %12
\frac{1}{2} \sigma^2  =  \int_0^t D dt  =  n T \frac{1}{T}  \int_0^T D dt 
\end{equation} 
Denoting the mean value of $D$ over the period $T$ by
\begin{equation} %13
\overline D  =  \frac{1}{T}  \int_0^T D dt 
\end{equation} 
and introducing time $t' = n T$ counted in new units $[T]$  
we obtain
\begin{equation} %14
f(x,t') = \frac{1}{2 \sqrt{ \pi \,  \overline D t'}} \ {\rm e}^{- \, \frac{x^2}{4 \overline D t' }}
\end{equation} 
We observe in more coarse time units the classical  brownian movement formula is recovered. 

\subsection{2 scale time homogenisation of  the brownian motion of stock prices }
To the analogous result we arrive applying more general method of asymptotic homogenisation, cf. 
[17, 18].
We introduce  two time variables $t$ and $\tau $ measured in different  scales, it is in different units of time. The  time $t$ is measured by a slow clock  and time $\tau$  by a fast (more accurate) clock. 
We have    
\begin{equation} %15
\tau = \displaystyle \frac{t}{\varepsilon}
\end{equation} 
where the scale parametr $ \varepsilon$ is positive ($ \varepsilon > 0$) and small.
For example, if
$[t]$ = day (the duration of a session $\equiv$ 6 hours) and   $[\tau]$= hour,  then 
$ \varepsilon = {\rm hour} / {\rm day}\approx  1 / 6$. 

Instead of $f(x,t)$ we write $f(x,t,\tau)$ and observe that   
$$
\frac{\partial f(x, t, \tau)}{\partial t} =   
\frac{\partial f(x, t, \tau)}{\partial t} +  \frac{\partial f(x, t, \tau)}{\partial \tau} \frac{1}{\varepsilon} 
$$
We assume an Ansatz
$$
f^{\varepsilon} = f^{(0)}(x, t, \tau) + 
 \varepsilon f^{(1)}(x,t, \tau) + \varepsilon^2 f^{(2)}(x,t, \tau) + \cdots  
$$
Then the diffusion equation (2) can be written in the form
{\renewcommand{\arraystretch}{2.3}
\begin{equation} %16
\begin{array}{l}
\displaystyle
\left(\frac{\partial}{\partial t} + \frac{1}{\varepsilon} \frac{\partial}{\partial \tau}\right) \
\left(f^{(0)}(x, t, \tau) + 
\varepsilon f^{(1)}(x,t, \tau) + \varepsilon^2 f^{(2)}(x,t, \tau) + \cdots  \right) \\
 =  
\displaystyle
D(\tau) \frac{\partial^2}{\partial x^2} \
\left(f^{(0)}(x, t, \tau) + \varepsilon f^{(1)}(x,t, \tau) + \varepsilon^2 f^{(2)}(x,t, \tau) + \cdots  \right)
\end{array}
\end{equation} 
}
We compare expressions at the same powers of $\varepsilon$, and   find consecutively: 
At $\varepsilon^{-1}$
$$
\frac{\partial f^{(0)}(x, t, \tau) }{\partial \tau} = 0 
$$
what means that $ f^{(0)} $ does not depend on the quick variable $\tau$
\begin{equation} %17
 f^{(0)} = f^{(0)}(x, t ) 
\end{equation} 
At $\varepsilon^0$ we have
\begin{equation} %18
\frac{\partial f^{(0)}}{\partial t} +  \frac{\partial f^{(1)}}{\partial \tau} =
D(\tau) \, \frac{\partial^2}{\partial x^2} f^{(0)}(x, t) 
\end{equation} 
We put 
\begin{equation} %19
f^{(1)} = \chi(\tau)  \frac{\partial f^{(0)}}{\partial t}
\end{equation} 
where $\chi(\tau)$ is a periodic function such that
\begin{equation} %20
\int_0^T  \chi(\tau)  d\tau = 0
\end{equation} 
After substitution (19) into (18) we get 
\begin{equation} %21
\frac{\partial f^{(0)}}{\partial t} \left(1 + \chi(\tau)\right)  =
D(\tau) \frac{\partial^2 f^{(0)}}{\partial x^2} 
\end{equation} 
Taking of mean with respect to variable $\tau$ over period $T$ gives
$$ 
\frac{\partial f^{(0)}}{\partial t}   =
\left(\frac {1}{T} \int_0^T  D(\tau) d\tau \right) \,
\frac{\partial^2 f^{(0)}}{\partial x^2} 
$$ 
or 
\begin{equation} %22
\frac{\partial f^{(0)}}{\partial t}   = \overline D \,
\frac{\partial^2 f^{(0)}}{\partial x^2} 
\end{equation} 
where definition (13) of the mean diffusion coefficient was used.
 The solution of the last equation with the initial condition $f(x, 0) =  \delta(x)$ is again given by (14), if only introduce  $t$ instead of $t'$, according  to the present meaning of time $t$ as a slow variable.

\section{Coefficient of diffusion dependent  on price change }
Consider Fick's first law with convolution, a more general  than (4),  
\begin{equation} %23
j (x, t) = - \int_{-\infty}^\infty  D(x-\xi) \frac{\partial f(\xi, t)}{\partial \xi} d \xi
\end{equation} 
Then instead of (1) we have the following equation of diffusion
\begin{equation}%24
\frac{\partial f}{\partial t} = 
\frac{\partial }{\partial x}\left(
\int_{-\infty}^\infty  D(x-\xi) \frac{\partial f(\xi, t)}{\partial \xi} d \xi \right)
\end{equation}
To both sides of the equation we apply the Fourier transformation and get
\begin{equation}%25
\frac{\partial \tilde f(k, t)}{\partial t} =   (i k)^2 {\tilde D}(k)  \tilde f (k, t)
\end{equation}
where
\begin{equation} %26
\tilde f (k, t) = \int_{-\infty}^\infty f(x, t) {\rm e}^{ikt} dx \quad {\rm and} \quad
\tilde D (k) = \int_{-\infty}^\infty D(x) {\rm e}^{ikt} dx 
\end{equation}
Solution of (25) reads
\begin{equation}%27
\tilde f (k, t) =  {\rm e}^{- (i k)^2 {\tilde D}(k)   t} 
\end{equation}
or
\begin{equation}%28
\tilde f (k, t) =  {\rm e}^{- \gamma  k^2} 
\end{equation}
where
\begin{equation}%29
 \gamma \equiv {\tilde D}(k)  t 
\end{equation}  
The function $\tilde f (k, t) $ in form (28) is known as the characteristic function of  Gauss stochastic process, cf. [19]. 

Assume that the trandorm of  diffusion coefficient $\tilde D$ is such that 
\begin{equation}%27
 \gamma = \gamma_0 k^{- \mu } 
\end{equation}  
where $\gamma_0$ depends linearly on $t$ but does not depend on $k$ while  $\mu$ is a positive constant. If 
\begin{equation}%28
 \alpha \equiv 2 -  \mu  
\end{equation} 
satisfies inequalities    
\begin{equation}%29
 0 < \alpha \le 2   
\end{equation} 
we deal with the L-stable process, cf. [19]. 

\section{Conclusions}

Above we tried to find a compromise between the classical view on finance  as a gaussian process and the modern view insisting on L\'evy form of stock price changes.
We have shown that:\\
1. In the case of periodically varying standard deviation of prices, the averaging over time period  restitutes gaussian character of the process.\\
2. Introduction of convolution in the diffusion equation may lead to the L\'evy distribution.

\section*{ References}

\small
\begin{enumerate}
\item %1.
L. Bachelier, % Th\'eorie de la sp\'eculation, (Th\`ese), 
{\it Annales scientifiques de l'Ecole Normale Sup\'erieure}, 3e s\'erie, {\bf  17}, pp 21-86,  Gauthier-Villars, Paris1900. 
Th\`ese soutenue le 29 mars 1900. 
R\'e\'edit\'e par Jacques Gabay, 1984, 1995: 
L. Bachelier,  {\it Th\'eorie de la sp\'eculation}. %,  Ed. Jacques Gabay, Paris 1995.\\

\item %2
R. Brown,  
%A brief account of microscopical observations made in the months of June, July, and August 1827, %on the particles contained in the pollen of plants; and on the general existence of active molecules in %organic and inorganic bodies, 
{\it The Philosophical Magazine} {\bf 4}, 161 (1828)

\item %3
A. Einstein,
%{\it  \"Uber die von molekularkinetischen Theorie der W\"arme geforderte Bewegung von in %ruhenden Fl\"ussigkeiten suspenddierten Teilchen, 
%On the movement of small particles suspended in stationary liquids required by the molecular-kinetic %theory of heat
%On the motion — required by the molecular kinetic theory of  heat—of  small particles suspended in %a stationary  liquid },   
{\it Annalen der Physik} {\bf 17}, 549 (1905). %549-560 

\item %4
A. Einstein, {\it The Collected Papers of Albert Einstein}, ed. John Stachel, vol. 2, The Swiss years,  Princeton Univ. Press, Princeton NJ 1989.

\item %5
M.  Smoluchowski, 
%Zarys teoryi kinetycznej ruchu Browna i roztwor\'ow m\c{e}tnych,
%Essai d'une th\'eorie cin\'etique du movement Brownien et des milieux troubles,
{\it Bulletin de l'Acad\'emie des Sciences de Cracovie}, 
Classe des Sciences math\'ematiques et naturelles, 
 N$^0$ 7, 577-602, 
%Imprimerie de l'Universit\'e, Cracovie, 
Juillet 1906.

\item %6
M. v. Smoluchowski,
% Zur kinetischen Theorie der Brownschen Molekularbewegung    und  der Suspensionen, 
%The kinetic theory of Brownian molecular motion and suspension, 
{\it Annalen  der Physik} {\bf  21}, 756 (1906). %756-780

\item %7
S. Chandrasekhar, M. Kac, R. Smoluchowski,  
{\it Marian  Smoluchowski, his life and scientific work}, ed. by R. S. Ingarden, PWN, Warszawa 1986

\item %8
S. Brush,  
%A history of random processes: Brownian movement from Brown to Perrin,  
{\it Archive for History of Exact Sciences}  {\bf 5},  1 (1968)

\item %9. 
F. Black and  M. Scholes, 
%The pricing of options and corporate liabilities,
{\it Journal of Political Economy} {\bf  81}(3),  637 (1973). % pp. 637-654. 

\item %10.
 J. P. Bouchaud, Y. Gefen, M. Potters, M. Wyart,
%Fluctuations and response in financial markets: the subtle nature of 'random' price changes,
{\it Quantitative Finance} {\bf  4}(2), 176 (2004).

\item %11
P. A. Samuelson, 
%Mathematics of Speculative Price 
{\it SIAM Review} {\bf  15}(1), 1 (1973). % (Jan., 1973) , pp. 1-42

\item %12
B. Mandelbrot, 
% The variation of certain speculative prices,
Journal of Business {\bf 36}, 394 (1963). %394-419

\item %13
B. B. Mandelbrot and J. R. Wallis, 
%Noah, Joseph, and operational hydrology, 
{\it Water Resources Research} {\bf  4}, 909  (1968) %pp. 909-918

\item %14
M. F. M. Osborne,  
%Periodic structure in the Brownian movement of stock prices, 
{\it Operations Research} {\bf 10}, 245 (1962). % (May-June 1962), pp. 245-279.

\item %15
A. I. Neishtadt,  T. V. Selezneva,  V. N. Tutubalin,  E. G. Uger, 
%Refinement of the L. Bachelier "theory of speculations",
{\it Obozrenie  Prikladnoi i Promyshlennoi  Matematiki} {\bf  9}(3), 525 (2002). %525-543

\item %16 
S. Chandrasekhar, 
%Stochastic problems in physics and astronomy,
{\it Reviews of Modern Physics}, {\bf 15}(1), 1 (1943). %1-89

\item %17
G. Sandri, 
%A new method of expansion in mathematical physics, 
{\it Nuovo Cimento} {\bf 36}(1), 67 (1965).

\item %18
E. Sanchez-Palencia, {\it Non-homogeneous media and vibration theory}, Lecture Notes in Physics, No. 127, Springer-Verlag, Berlin 1980. 

\item %19
R. N. Mantegna, H. E. Stanley, {\it An introduction to econophysics. Correlations and complexity in finance}, Cambridge University Press 2000.

\end{enumerate}

\end{document}